# Decoupling of a Current-Biased Intrinsic Josephson Junction from its Environment

P. A. Warburton and A. R. Kuzhakhmetov

Dept. of Electronic and Electrical Engineering, University College London, Torrington Place, London, WC1E 7JE, UK.

G. Burnell and M. G. Blamire

Dept. of Materials Science, University of Cambridge, Pembroke St., Cambridge, CB2 3QZ, UK.

H. Schneidewind

IPHT-Jena, 07745 Jena, Germany

ABSTRACT

We have observed a dissipative phase diffusion branch in arrays of hysteretic high-$T_c$ intrinsic Josephson junctions. By comparing the data with a thermal activation model we extract the impedance seen by the junction in which phase diffusion is occurring. At the plasma frequency this junction is isolated from its environment and it sees its own large (~ k$\Omega$) impedance. Our results suggest that stacks of Josephson junctions may be used for isolation purposes in the development of a solid state quantum computer.

PACS: 85.25.Cp, 74.81.Fa, 74.72.Jt, 03.67.-a

I. INTRODUCTION

The Josephson junction [1-6] and the Josephson junction array [7-9] are ideal systems on which to perform experiments on quantum phenomena such as macroscopic quantum coherence, single-electron tunnelling and quantum critical phenomena. The Josephson junction is described by two conjugate macroscopic variables, namely the phase difference, $\varphi$ across the junction and the charge on it, $q$. The well-defined variable can be either $\varphi$ or $q$ depending upon the ratio of the Josephson energy, $E_J$, to the charging energy, $E_c$. This ratio can be controlled both during fabrication (by changing the junction area) and *in situ* (by applying a magnetic field).

In such experiments it is essential that the quantum coherence is not destroyed by decoherent interactions with the environment. The fact that quantum phenomena are generally observed in small Josephson junctions with large normal-state resistance, $R_N$, thus poses the following problem: the impedance seen by the junction (the dissipative part of which determines the coherence lifetime) is not $R_N$, but the shunting impedance of the lines which make connections to it, which at the frequencies of interest (~ GHz) is less than $R_N$. What is required therefore is a small Josephson junction attached to current bias leads which is isolated from the environment so that it sees it own large resistance.

Recently it has been shown that it is possible to control the impedance seen by a low-$T_c$ junction by integrating a low-value resistor in parallel with it [10]. Here we report our experiments on a novel approach to enable a junction to see its own *large* impedance even when attached to current biasing circuitry. We have used high-$T_c$ intrinsic Josephson junctions (IJJ's) [11]. IJJ's are formed in series arrays of $N$ junctions, the spatial periodicity of the array being ~ 1.7 nm. Since the IJJ's are so closely spaced, the impedance seen by one oscillating junction is that of the other $N - 1$ in series. At the plasma frequency, $f_p$, the impedance of these $N - 1$ junctions is maximised, decoupling the oscillating junction from its environment and allowing it to see its own resistance. This approach is similar to that used with low $T_c$ junctions by Delsing *et al.* [7], although in this case the spatial periodicity of the junctions is two orders of magnitude greater than for IJJ's.

II. EXPERIMENTAL

We have observed a dissipative phase-diffusion branch (*i.e.* a non-zero-voltage branch of the current-voltage characteristics at current bias values less than the critical current) in hysteretic IJJ arrays. Since the spacing between adjacent copper-oxide double planes (~ 1.5 nm) in an anisotropic high temperature superconductor is greater than the c-axis coherence length, such planes are Josephson coupled. Hence for transport currents normal to the planes, a single crystal of such a superconductor acts as a series array of IJJ's [11]. We have created such an array by patterning (using a gallium focussed ion beam) a bridge of width $w$ = 500 nm and length $L$ = 3 μm in a $Tl_2Ba_2CaCu_2O_8$ (TBCCO) thin film of thickness $b$ = 240 ± 60 nm grown on a 20°-vicinal $LaAlO_3$ substrate. The bridge structure is shown in the inset to figure 1 (b). Epitaxial growth of the TBCCO film on the vicinal substrate results in a 20° mis-alignment of the copper-oxide double planes with respect to the substrate surface. This fabrication strategy allows us to measure c-axis coupling with a transport current which is parallel to the substrate. Indeed, since the critical current density in TBCCO is strongly anisotropic, the voltage drop across the bridge when it is current biased is due to dissipation resulting from c-axis coupling. This, as we will show below, is due to slip of the Josephson phase difference between adjacent copper-oxide double planes.

The number of junctions in the bridge is given by

$$N \approx \frac{L\sin 20° - b\cos 20°}{x_c} \tag{1}$$

where $x_c$ is the periodicity of the real space crystal lattice in the c-direction. Here the $b\cos 20°$ term accounts for the junctions at either end of the bridge which are shorted by in-plane transport currents. The bridge reported here therefore contains 460±50 IJJ's, the uncertainty arising due to the fact that the ends of the bridge are not well defined with respect to the sloping copper-oxide double planes. The TBCCO epitaxy is confirmed both by x-ray diffraction and transmission electron micrography [12]. Four terminal contacts are made to the bridges along TBCCO lines patterned in the same film. Details of film growth [13] and device fabrication [12] are found elsewhere. Transport measurements are made using a low-frequency (~10 Hz) linear current bias sweep. The sample is mounted at the end of a probe immersed in helium vapour in a dewar. The sample stage is shielded by a mu-metal cap. Coaxial lines along the length of the probe are coupled to the room-temperature current source and low noise amplifier *via LC* filters with a cut-off frequency of 200 Hz.

The current-voltage characteristics at 4.2 K are shown in figure 1. The supercurrent branch is close to zero voltage and extends up to a bias current of 5.50 µA. This supercurrent branch will be analysed in detail in section III. The multi-branched structure at voltages greater than a few mV is similar to that observed for both IJJ's fabricated from single crystals and mis-aligned TBCCO thin film IJJ's of larger area that we have previously reported [12]. Here each branch corresponds to the switching of an additional IJJ to its quasiparticle branch. Hysteretic and discontinuous features within each branch arise due to resonance of the Josephson oscillations with c-axis phonon modes in $Tl_2Ba_2CaCu_2O_8$ [14, 15]. The dependence of the critical current, $I_c$, (measured using an arbitrary voltage criterion of 100 μV) upon an in-plane aligned magnetic field shows a minimum at 2.5 T [16]. Based on the geometry of a single junction in this bridge we estimate that the field required to insert a single flux quantum per junction is 1.7 ± 0.6 T. While the comparison of these two figures is not perfect it is clear that the Josephson effects we observe are not due to extrinsic phenomena such as coupling across a grain boundary or across the whole 3μm length of the bridge. We conclude that the transport mechanism is intrinsic Josephson coupling between adjacent copper-oxide double planes.

III. ANALYSIS

Inspection of the low-voltage part of the characteristics reveals that the supercurrent branch in zero magnetic field is *not at zero voltage*, in contrast with IJJ's of larger area [17]. As shown in figures 1 (b) and 2, there is dissipation at currents lower than $I_c$ in the sub-micron IJJ's reported here. We now consider the possibility that this dissipation is due to thermally activated phase diffusion. The dynamics of the phase difference, $\varphi$, in a current-biased Josephson junction are the same as those of a massive particle moving in a periodic "washboard" potential

$$U(\varphi) = -E_J\left(\cos\varphi + \frac{I}{I_c}\varphi\right), \qquad (2)$$

where $E_J = \hbar I_c/(2e)$, $I$ is the bias current, $2\pi\hbar$ is Planck's constant and $e$ is the magnitude of the electronic charge [18]. Escape from a local minimum is achieved by thermal activation over a potential barrier of height

$$\Delta U \approx 2E_J\left(1 - \frac{I}{I_c}\right)^{3/2}. \qquad (3)$$

For an underdamped junction the particle then has enough inertia to traverse subsequent minima and the phase will advance $2\pi$ for every traversed minimum. If the junction is overdamped the particle will, after a few traversals, be retrapped. This is known as phase diffusion, and leads to a voltage proportional to the average rate of change of the phase difference. For junctions of sufficiently small $I_c$, this voltage can be measured at $I < I_c$. A full theoretical analysis of this was published by Ivanchenko and Zil'berman [19]. They found that the supercurrent reaches a peak value at low voltage before decreasing with increasing voltage.

The phase diffusion voltage, $V$, exponentially depends upon $I_c$. Inspection of the current-voltage characteristics over the full bias range [12] shows, in common with other thin film IJJ's, a broad spread in the $I_c$'s of the $N$ junctions in the bridge. Our analysis is therefore based upon the assumption that all the voltage is due to phase diffusion in the single IJJ with the lowest $I_c$. This junction we call the source junction. The other $N$-1 junctions are not dissipative and we call them the load junctions.

The analysis of Ivanchenko and Zil'berman [19] can be very well approximated by a simple thermal activation model, which breaks down only near the supercurrent peak and at higher voltages. (Voltages beyond the supercurrent peak are in any case inaccessible in our experiment since our bridge is current-biased.) For a junction of resistance $R_N$ connected to leads of infinite impedance this model yields

$$V = I_c R_N \exp\left(\frac{-\Delta U}{kT}\right), \quad (4)$$

where $k$ is Boltzmann's constant, $T$ is the absolute temperature and $\Delta U$ is given by equation (3). Here we neglect activation in the "uphill" direction, this being valid provided that

$$\exp\left(\frac{\pi\hbar I}{kTe}\right) \gg 1. \quad (5)$$

For the general case of a junction connected to a load of frequency-dependent complex impedance $Z(\omega)$, equation (4) is modified to

$$|V| = I_c |Z(\omega)| \exp\left(\frac{-\Delta U}{kT}\right) \quad (6)$$

If $Z$ were independent of frequency the experimental approach would be to plot log($V$) at various values of current bias as a function of 1/$T$ in order to extract the current dependence of the activation energy $\Delta U$. Since, however, $Z$ is strongly frequency dependent (as we will show below), we must extract data at a fixed Josephson frequency – *i.e.* at a fixed voltage. Inverting equation (6) and substituting for $\Delta U$ from equation (3) we obtain

$$1 - \frac{I}{I_c} = \left\{ \ln\left(\frac{I_c |Z(\omega)|}{|V|}\right) \frac{k}{2E_J} T \right\}^{2/3} \quad (7).$$

Hence plots of the current, $I$, versus $T^{2/3}$ at fixed voltage should extrapolate to $I = I_c$ at $T$=0. The slope allows us to extract $|Z(\omega)|$ at each voltage – *i.e.* at each frequency. Here we have assumed that $I_c$ is independent of $T$. It has been shown [11, 20] that IJJ's follow the Ambegaokar-Baratoff relationship [21] for the temperature-dependence of $I_c$. We therefore restrict our analysis to temperatures below $T_c/2$, in which regime $I_c$ is essentially constant.

The temperature dependence of the current at three fixed voltages is shown in figure 3. These data are obtained by off-line analysis of the experimental data which were taken under a current bias. For $T > 14$ K the data are linear, with an extrapolated critical current of 7.2 ± 0.1 µA. The deviation from linearity at low $T$ is due to the fact that the thermal activation model breaks down near the supercurrent peak [22]. We have observed similar behaviour in another bridge fabricated using the same technique but in a different TBCCO film. The $I_c$ for this bridge was ~ 70 nA, confirming that our thermal activation model is valid over a wide range of $I_c$. For this latter bridge, however, the inequality (5) restricts us to a very narrow temperature range, preventing us from making a more quantitative analysis.

The goodness of the fit to the data in figure 3 supports our suggestion that the dissipation mechanism in the supercurrent branch is phase diffusion. From the slope of the linear part of curves like those in figure 3 we may now extract (by use of equation (7)) the magnitude of the impedance seen by the source junction. This we plot as a function of frequency in figure 4, the frequency being obtained from the voltage by use of the Josephson relationship. The form of the curve is suggestive of an overdamped parallel *LCR* circuit, with the peak occurring at the resonant frequency.

Such a resonance can arise from artefacts such as geometrical resonances. It is however likely that the load seen by the source junction is the impedance of the *N*-1 junctions which are in close proximity to it. Here we will use the "RSJ" model [18] of a Josephson junction (as shown in figure 5(a)) for both the source junction and each load junction. Since phase diffusion occurs in the source junction it must be overdamped, so we take the capacitance of the source junction to be zero. We retain the possibility that the load junction capacitances are non-zero, as suggested by the hysteresis of the current-voltage characteristics. The a.c. equivalent circuit is shown in figure 5(b). Hence the net impedance seen by the source junction is

$$Z_{net} = \left( \frac{1}{R_N} + \frac{1}{(N-1)Z_{load}} \right)^{-1} \qquad (8),$$

where $Z_{load}$ is the impedance of a single load junction, given by

$$Z_{load} = \left( \frac{1}{R_N} + i \left( 2\pi fC - \frac{1}{2\pi fL} \right) \right)^{-1} \tag{9}.$$

Here $C$ is the single junction load capacitance, $L$ is the Josephson inductance of a single load junction and $f$ is the frequency. Hence it can be seen that in this model $Z_{net}$ is inductive at frequencies below the plasma frequency, $f_p = (LC)^{-0.5}/ 2\pi$, and capacitive at $f > f_p$. At the plasma frequency, $Z_{load} = R_N$, and the total resistance of the load junctions is $N$-1 times larger than the resistance of the source junction. Hence the impedance seen by the source is its own resistance.

In figure 4 we show the frequency dependence of the magnitude of $Z_{net}$ for our model circuit, given by equations (8) and (9), with $N$ = 460 as *per* the experiment. The free parameters are $L$, $C$ and $R_N$. $C$ is determined by fitting to the high frequency data and $R_N$ from the peak at resonance. $L$ is then found from the extracted value of $C$ and the resonant frequency. We then compare these extracted values with those which we expect for our IJJ's.

From the fit in figure 4 we find $C$ = 28 ± 3.6 fF. By treating each IJJ as a parallel plate capacitor and taking an estimate of 10 for the dielectric constant we predict the single junction capacitance to be 22 fF, in reasonable agreement with our extracted experimental value. The extracted source junction resistance is $R_N$ = 7.8 ± 0.5 kΩ, yielding a low temperature $I_cR_N$ value of 56 ± 4 mV. By using the Ambegaokar – Baratoff expression we obtain a gap energy of 36 ± 3 meV in TBCCO [23]. This is comparable with measurements obtained by tunnelling of between 25 and 30 meV [24, 25]. From the measured resonant frequency of 220 ± 40 GHz we find $L$ = 18.7 ± 7.3 pH [26]. For bias currents less than $I_c$ / 2 the inductance of a Josephson junction is $\Phi_0/(2\pi I_c)$, where $\Phi_0$ is the flux quantum. We therefore estimate an average $I_c$ of 18 ± 7 µA for the load junctions. Since the source junction is that with the lowest $I_c$ (equal to 7.2 ± 0.1 µA here) and we measure $I_c$ spreads on the order of 2 to 3 times the mean, this extracted average $I_c$ is consistent with the measured source junction $I_c$. Hence we find that the measured load impedance seen by the source junction is consistent with the impedance of N-1 = 460 junctions in series (away from $f_p$ = 220 GHz) and the resistance of the single source junction (near $f_p$).

From our extracted values of L, C and $R_N$ we infer a d.c. quality factor $Q = 2\pi f_p R_N C$ of 300 consistent with the underdamped (*i.e.* hysteretic) d. c. current – voltage characteristics. Our measurement of $f_p$ is a factor ~3.5 lower than that measured by far-infrared spectroscopy [27]. Grabert [28] has shown that in overdamped junctions $f_p$ is reduced by a factor $(1+(2Q)^{-2})^{1/2} - (2Q)^{-1}$. This suggests that at our measured $f_p$ = 220 GHz the quality factor is 0.31. This is consistent with the observation of phase diffusion for which it is necessary that Q<1 at $f_p$. We do not presently understand why one IJJ in the bridge (*i.e.* the source junction) appears to be more heavily damped at $f_p$ than the others (*i.e.* the load junctions), yielding the equivalent circuit shown in figure 5 (b). We speculate that this is due to variations in the microstructure of the film at the unit cell level, consistent with our observations [12] that the d.c. hysteresis appreciably varies from device to device even within the same sample.

IV. SUMMARY

We have observed phase diffusion in one-dimensional arrays of N ~ 460 intrinsic Josephson junctions. In general the impedance seen by the source junction in which phase diffusion occurs is the reactive impedance of the N – 1 load junctions. Near the plasma frequency (~ 220 GHz), however, the load impedance exceeds the resistance of the source junction, and the source junction *sees its own impedance* of 7.8 kΩ. The source junction has been isolated from its environment by the large series impedance of the other junctions near the plasma frequency. The ability to isolate junctions from their environment is essential in the development of long-lived coherent states on which to perform quantum computing operations. While the impedance we measure here is too low for such experiments and dissipation (in the form of phase diffusion) is present, both these issues can be resolved in principle by reducing the junction area. Our results suggest that closely-spaced stacks of small Josephson junctions may be suitable for isolation of solid-state qubits from their environment.

This work was supported by the UK EPSRC.

**FIGURE CAPTIONS**

Figure 1: (a) Current-voltage characteristics at 4.2 K in zero magnetic field showing the supercurrent branch and ten of the quasiparticle branches (only data below 100 mV are shown). (b) shows the supercurrent branch only at a larger voltage scale. The inset shows the schematic of our device geometry (not to scale). The orientation of the copper-oxide double planes is indicated by the slanting lines in the bridge. The dimensions are $b$ = 240 nm, $L$ = 3 µm, $w$ = 500 nm. $I$ is the bias current.

Figure 2: Current-voltage characteristics of the bridge in zero magnetic field. Only the supercurrent branch is shown. The voltage is shown on a logarithmic scale. The temperature is 4.2 K (lowest curve), 7 K, 9 K, then in 2 K intervals up to 39 K (uppermost curve).

Figure 3: Temperature dependence of the current in zero magnetic field measured at voltages of 0.5 mV (squares), 0.75 mV (circles) and 1 mV (diamonds). The lines are linear fits to the data at temperatures above 14 K.

Figure 4: Frequency dependence of the magnitude of the impedance seen by the source junction. The data (diamonds) are extracted from the gradients of the linear fits to data such as those shown in figure 3 and by using equation (7). The line shows the magnitude of the impedance seen by the junction in the model circuit of figure 5 (b), given by equations (8) and (9). The fitting parameters are $C$ = 28 fF, $L$ = 18.7 pH and $R_N$ = 7.8 kΩ.

Figure 5: (a) the "RSJ" model of a single Josephson junction. (b) a.c. equivalent circuit of the overdamped source junction in parallel with the ($N$-1) load junctions.

FIGURE 1

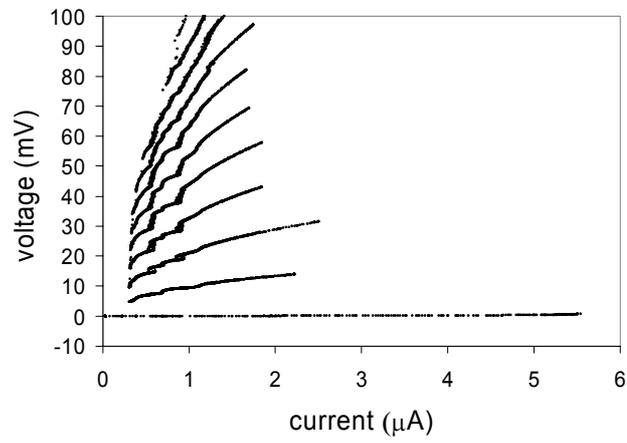

(a)

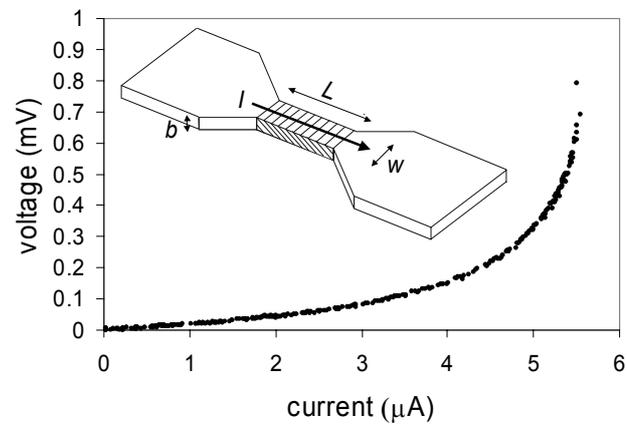

(b)

FIGURE 2

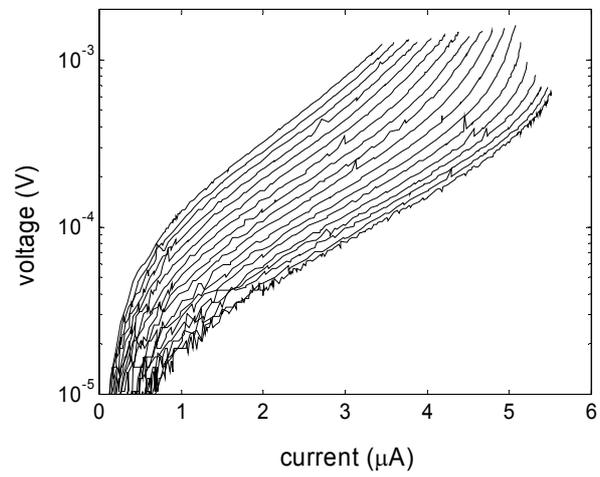

FIGURE 3

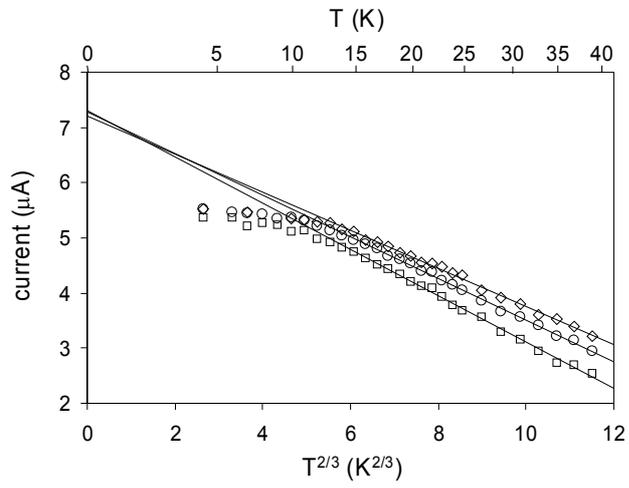

FIGURE 4

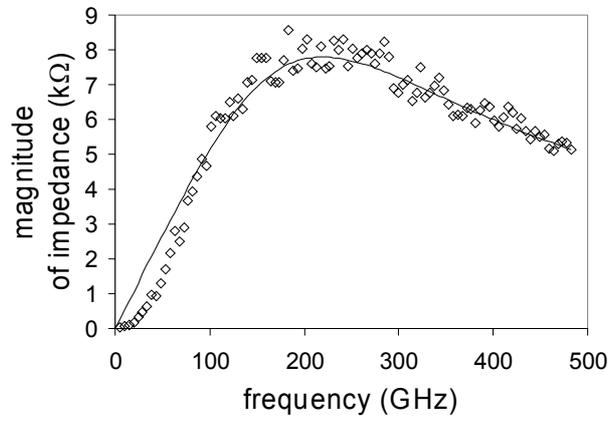

Warburton *et al.* "Decoupling of a current-biased intrinsic Josephson junction..."

FIGURE 5

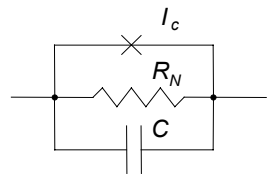

(a)

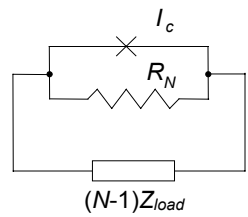

(b)